\documentclass[a4paper,10pt]{edpsci} 
\usepackage{amsmath}
\usepackage{graphicx}
\usepackage{epstopdf}
\usepackage[numbers]{natbib}
\usepackage{hyperref}
\usepackage{url}
\hypersetup{colorlinks=true,citecolor=blue,urlcolor=blue,linkcolor=blue}

\usepackage{lineno}

\usepackage{multirow}
\journalname{Acta Acustica}

\articletype{Research article}

\begin{document}

\title{Violin ``Playing-In'': Disentangling Physical Change from Player Adaptation via Physical Measurements}

\titlerunning{Violin ``Playing-In'': Physical Measurements}


\author{Hugo Pauget Ballesteros\inst{1}\correspondingauthor{\email{hugo.pauget@dalembert.upmc.fr}} 
\and
Philippe Lalitte\inst{2}
\and
Vincent Lostanlen\inst{3}
\and
Claudia Fritz\inst{1}}

\authorrunning{H. Pauget Ballesteros et al.}


\institute{Sorbonne Université, CNRS, Institut Jean Le Rond d'Alembert, F-75005 Paris, France\and
Sorbonne Université, CNRS, Ministère de la Culture, Bibliothèque nationale de France, Institut de Recherche en Musicologie, IReMus, F-75018 Paris, France\and
Nantes Université, Centrale Nantes, CNRS, LS2N, UMR 6004, F-44000 Nantes, France}

\abstract{It is a widespread belief among musicians that a violin's sound ``opens up'' or improves through regular playing. However, physical evidence for this ``playing-in'' effect remains elusive. This study revisited the phenomenon by testing two hypotheses: (1) the instrument undergoes physical evolution, or (2) the player undergoes behavioral adaptation. We conducted a longitudinal study centered on a seldom-played test violin played daily by a professional soloist for six months, alongside two stored control violins and a control group of ten violinists (N=10). All three violins had been rarely played prior to the experiment. Assessments performed at the beginning (``Before'' phase) and end (``After'' phase) of the period included input admittance measurements via laser vibrometry, standardized sound recordings, and acquisition of the soloist's bowing gestures via motion capture. Results revealed no acoustical evolution in the played violin exceeding the environmental drift observed in the control instruments. Furthermore, kinematic analysis showed no significant drift in the soloist's bowing strategy (bow position, velocity, contact point). These findings suggest that the ``playing-in'' phenomenon is driven neither by macroscopic physical changes in the instrument nor by a fundamental reorganization of the player's technique.}



\maketitle

\section{Introduction}
Luthiers and violinists often claim that a new or dormant instrument requires a period of musical practice (``playing-in'') before reaching its full potential (``opening up''). For instance, when describing his relationship with the instrument recently loaned to him, the soloist Elias David Moncado says:

\begin{quotation} \small\it
    I would name Szigeti as one of my sound idols, alongside David Oistrakh, Isaac Stern, and Jascha Heifetz, and, in many ways, I can sense the influence and the impact it had on the violin; it is as if the soul of Szigeti continues to live on.

    I immediately sensed an incredible sound potential, and as I spend my time intensely performing and practicing on it, I continue to discover new sound colors which I didn’t know existed. The instrument ``opens up'' even more and unveils a new dimension of possibilities.
    
    \hspace{1em plus 1fill}--- Elias David Moncado, on The Violin Channel
\end{quotation}

Moncado explicitly refers to the concept of ``opening-up'', reached after a certain time of practice, and even feels that a previous owner (Joseph Szigeti) had physically imprinted his sound on the violin. Other string players have noted that this experience is so universally shared that it transcends the need for a scientific explanation:

\begin{quotation} \small\it
    [T]here are many theories which attempt to explain [this effect] and there is validity behind the science. Plate vibrations, bridge and soundpost settling into the plates, wood drying so sap pockets become air chambers, etc. But we honestly don't know why [violins] require breaking-in. No matter the reasoning it's a no-brainer we string players accept without precise explanation and trust simply from pure experience.
    
    \hspace{1em plus 1fill}--- Fiddleheads.com
\end{quotation}

The strength and consistency of such testimonies have prompted scientific efforts to determine whether the perceived ``playing-in'' effect can be objectively measured and explained.  Early research often reported measurable vibro-acoustic changes or improved subjective ratings \cite{bissinger1995, hutchins1998, ling1997}; however, these studies frequently lacked non-vibrated control groups or rigorous blinding protocols, making it difficult to isolate vibrational effects from environmental fluctuations or observer bias. While some material studies suggest that sustained vibration may alter the rheological properties of spruce, such effects often appear reversible or dependent on extreme humidity conditions \cite{hunt1996}. In contrast, recent studies employing high experimental rigor—including ``twin-instrument'' designs, humidity tracking, and blind testing—have consistently failed to detect acoustical or perceptual changes that exceed environmental noise \cite{grogan2003, inta2005, piacsek2024b}. This creates a notable paradox: despite a growing body of ``null results'' in the literature, over 70\% of the community of luthiers and musicians remains convinced of the effect's material reality \cite{piacsek2025}. Finally, we face a methodological dilemma: excluding the musician removes the ecological validity of the excitation, but including the musician introduces human variability and motor learning, which have never been simultaneously monitored in this context.

Understanding the acoustical bases of playing-in has significant implications for instrument valuation and performance practice. On one hand, if the effect is physical, it challenges current models of wood mechanics and aging.
On the other hand, if it is behavioral, it suggests that the celebrated ``opening up'' is actually a manifestation of the musician's adaptability; that is, a sensorimotor optimization rather than a material evolution. 

This study addresses this issue by testing two hypotheses concerning the mechanism underlying the perceived playing-in effect. The first hypothesis is that the effect arises primarily from instrumental change: sustained excitation alters the violin’s vibro-acoustic characteristics, leading to systematic modifications in its radiated sound. The second and novel hypothesis is that the effect arises primarily from sensorimotor learning: with increasing familiarity, the violinist unconsciously adapts bowing gestures (e.g., bow force, speed, contact point, timing) to optimize the instrument’s acoustic output, thereby producing an apparent ``playing-in'' effect even if the instrument’s physical properties remain unchanged.

To disentangle these factors, we conducted a controlled longitudinal study over six months involving a dedicated ``Test Violinist'' and a specific ``Test Violin'', compared with a Control Group of ten violinists and two stored Control Violins. All three violins had been rarely played prior to the experiment. By combining physical measures on the three instruments alone (bridge mobility) or in interaction with  players (sound recording for the 11 players and gestural recordings only for the Test Violinist), we aimed to determine whether the ``opening up'' of a violin is a material reality or a motor adaptation.

The detailed design (protocol, participants, violins) as well as data collection and processing are described in Section \ref{sec:m&ms}. The results are presented in Section \ref{sec:results} and discussed in Section \ref{sec:d&c}.

\section{Materials and methods}\label{sec:m&ms}

\subsection{Study Design and Timeline}
The experiment followed a longitudinal protocol designed to disentangle physical instrument evolution from player adaptation. The study employed a double-control design comparing a ``played'' condition against a ``stored'' condition (by using one Test Violin and two Control Violins, all three having been rarely played prior to the experiment), and a ``Test Violinist'' against a ``Control Group.'' Measurements were conducted before and after an intensive playing period during which the Test Violinist played the Test Violin daily; the duration of this phase was not fixed but lasted until the Test Violinist reported that the instrument had reached the state of having ``opened up'' (which occurred at approximately 6 months).

The protocol was thus divided into three phases: a ``playing-in'' phase during which the Test Violinist played the
Test Violin daily (with minor exceptions during concert tours) while the Control Violins remained in storage; this phase was preceded by a phase called ``Before'' and succeeded by a phase called ``After''. The duration of the ``playing-in'' phase was not fixed in advance; rather, it concluded when the Test Violinist reported that the instrument had reached the state of having ``opened up'' (which occurred at approximately 6 months). 


\subsection{Participants}
The cohort consisted of 11 professional and semi-professional violinists. One musician, a professional soloist and violin professor, was designated as the Test Violinist and performed the daily playing intervention. An established violin maker stated that this violinist was ``the perfect candidate for opening a violin''. The remaining 10 participants served as a Control Group; this group consisted of professional and semi-professional violinists who were naive to the specific goals of the study and the roles of the instruments, participating solely in the measurement sessions.

\subsection{Violins}
Three modern violins (Klimke 2012, Levaggi 2012, Stoppani 2018) were selected for the study based on their history of being rarely played in the years preceding the experiment. The Test Violin (Klimke) was selected by the Test Violinist based on its perceived potential to ``open up'' and improve through playing. The first Control Violin (Levaggi) was selected for its high similarity in construction and varnish to the Test Violin; it was even crafted the same year, thus serving as an ideal  reference. The second Control Violin (Stoppani) served as a secondary reference.

While the Test Violin was played extensively, the Control Violins were kept in the instrumental collection of the \emph{Conservatoire National Supérieur de Musique et de Danse de Paris} (CNSMDP). This location provided a stable, instrument-friendly storage environment comparable to a professional luthier's workshop. To ensure consistent conditions, all three violins were adjusted by a luthier prior to the study, and equipped with identical strings and chinrests. The strings of the Test Violin were changed after the playing period.

\subsection{Measurements}
A set of metrics was acquired to monitor the evolution of the instruments (bridge admittance) and the player-instrument system (sound recording for the 11 players and bowing parameter recordings only for the Test Violinist).

\subsubsection{Bridge mobility}
Input admittance was measured at the bridge using a Laser Doppler Vibrometer (Polytec OFV-505/OFV-3001) and an instrumented impact hammer (PCB 086E80). Violins were placed horizontally on a block of soft foam resting on a heavy rigid table, and the strings were damped to minimize string resonances in the measurement. Excitation was applied to the bridge on the G-string side, and velocity was measured at the E-string side.

\subsubsection{Sound}
Recordings were acquired in a large dry room (approx. \(100\text{ m}^{2}\)) with low reverberation. Signals were captured using a stereo pair of DPA 4010 omnidirectional microphones positioned above the player. To ensure longitudinal repeatability, microphone gain was fixed via the audio interface (RME Fireface) across both phases, and player/microphone positions were standardized using floor markings.

The recording protocol for each violin consisted of a 5-10 minute familiarization period followed by the performance of standardized excerpts from the repertoire (scales and melodic phrases).


\subsubsection{Kinematic data}
\begin{figure}
\centering
\includegraphics[width=0.4\textwidth]{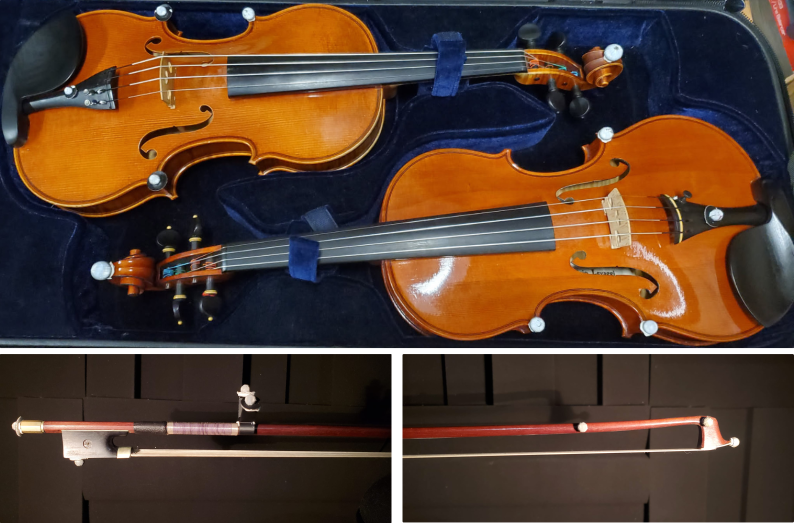}
\caption{Markers placed at different points on the violins and the bow.} \label{fig:methods.violin-markers}
\end{figure}

Kinematic data of the Test Violinist were acquired using motion capture via an OptiTrack Motive system comprising 12 cameras operating at 120 Hz. Passive reflective markers were attached to the Test Violin and their Personal Violin, as well as their bow (see Figure \ref{fig:methods.violin-markers}).

Due to the physical demands of the recording protocol (numerous repetitions), we collected data only on the Test Violin and the Test Violinist's Personal Violin. The Personal Violin was chosen as the reference standard: given the player's expertise and familiarity with their own instrument, it offers the most robust baseline for assessing stability and detecting potential adaptations to the Test Violin.

The procedure mirrored the audio recording protocol: the violinist performed the standardized repertoire after a warm-up period. 

\subsection{Data analysis}

\subsubsection{Bridge mobility processing}
For each measurement point, the Frequency Response Function (FRF) was estimated using the \(H_{1}\) estimator based on 5 consecutive impacts:

\[H_{1}(f) = \frac{\langle G_{xy,i}(f)\rangle}{\langle G_{xx,i}(f)\rangle}\]

where \(G_{xy,i}\) is the cross-power spectral density between the \(i\)-th impact force and velocity, \(G_{xx,i}\) is the auto-power spectral density of the force, and \(\langle \cdot \rangle\) denotes the average over the 5 impacts.

To quantify repeatability, measurements were performed three times for each phase. Due to the high reproducibility of laser vibrometry measurements \cite{piacsek2024c}, this sample size was considered sufficient. The representative amplitude curve for each phase was calculated as the Root Mean Square (RMS) value of the linear magnitudes:

\[|H|_{\text{phase}}(f) = \sqrt{\frac{1}{3}\sum_{k = 1}^{3}|H_{1,k}(f)|^{2}}\]

Given the limited sample size (\(N = 3\) repetitions for each phase), standard statistical confidence intervals were deemed inappropriate. Instead, measurement variability is represented by the min-max envelope of the recorded magnitudes. For each frequency bin, this interval is defined by the minimum and maximum dB values observed across the three independent measurement blocks.

\subsubsection{Audio Feature Extraction}
The Long-Term Average Spectrum (LTAS) was computed over the entire duration of the excerpt using Welch's method (Hann window, length 2048 samples, hop size 512 samples) at the native sampling rate of 48 kHz. To minimize the influence of outliers, the median was employed to aggregate the spectral frames. 

While the full bandwidth was analyzed, the results presented in this study focus primarily on the range up to 5 kHz. Previous work \cite{ballesteros2025} demonstrated that this frequency band contains the spectral features necessary for high-accuracy violin identification, suggesting that any significant physical evolution of the instrument's ``voice'' would likely be observable within this range.

\subsubsection{Bowing parameters extraction and processing}
\begin{figure}
\centering
\includegraphics[width=0.4\textwidth]{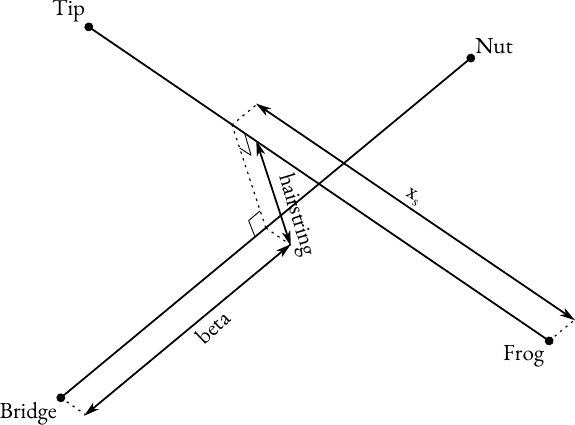}
\caption{Extracted bowing parameters.}\label{fig:methods.mocap-desc}
\end{figure}

Extracted features include bow position and velocity, bow-bridge distance, and the bow-string distance (see Figure \ref{fig:methods.mocap-desc}). The computation of these features follows \cite{salvadorcastrillo2024}, where it is described in detail. Notably, the bow-string distance (\textit{hairstring}) is computed as the signed distance between the bow hair and the string. A positive value indicates the bow is hovering above the string, zero indicates resting contact, and a negative value indicates the bow hair and string are actively bending under applied force.

To enable a direct comparison across repeated performances, it is necessary to account for the natural temporal variability inherent in human musical performance. The time-series data for each excerpt were temporally aligned using Dynamic Time Warping (DTW). The alignment cost was calculated based on the longitudinal bow speed (\(v_{s}\)), defined as the speed of the bow relative to the string. This parameter was selected as the alignment reference because its extrema and zero-crossings robustly encode the rhythmical structure of bowing gestures, particularly bow changes. For each musical excerpt, a reference take was selected from the ``Before'' phase, and all subsequent recordings were warped to match its temporal timeline.

\section{Results}\label{sec:results}

\subsection{Bridge mobility}
\begin{figure*}
\centering
\includegraphics[]{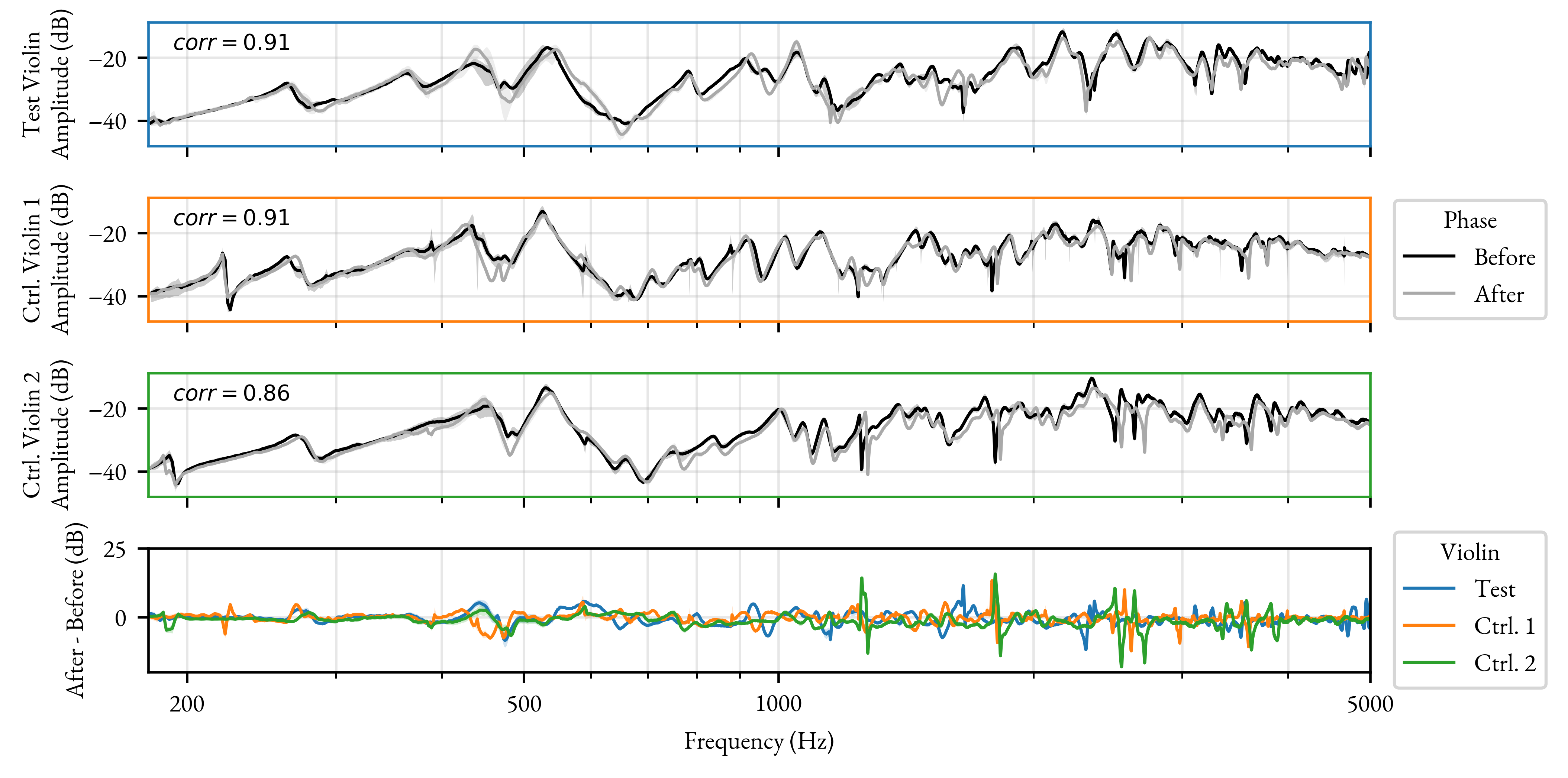}
\caption{Input admittance measured at the bridge for the Test Violin and the two Control Violins. Black curves indicate the ``Before'' phase and gray curves indicate the ``After'' phase; the Pearson correlation coefficient between these two states is provided in the top-left corner of each panel. The bottom panel displays the magnitude difference between the ``Before'' and the ``After'' phases for each instrument. Shaded areas represent the measurement uncertainty (minimum and maximum values over 3 repetitions).}\label{fig:admittances}
\end{figure*}

Figure \ref{fig:admittances} presents the measured bridge mobilities for the Test Violin  and the two Control Violins. The top three panels show the frequency response functions for each instrument, while the bottom panel shows the spectral difference between the two phases for the three violins.

Visually, the frequency response functions exhibit high stability, with the ``After'' curves (gray) largely superimposing onto the ``Before'' curves (black) for all instruments. This visual stability is quantitatively confirmed by the high Pearson correlation coefficients computed between the two phases.

Regarding the spectral differences, while localized fluctuations occurred, it is crucial to note that the deviations for the Test Violin are of the same order of magnitude as those observed for the Control Violins. This observation is further supported by the correlation coefficients, which remain highly comparable across all three instruments. The spectral changes for the played instrument do not exceed the baseline variability for the unplayed instruments, suggesting that the measured variations are attributable to environmental factors or measurement reproducibility rather than a specific ``playing-in'' effect.

\subsection{Sound features}

\begin{figure*}
\centering
\includegraphics[]{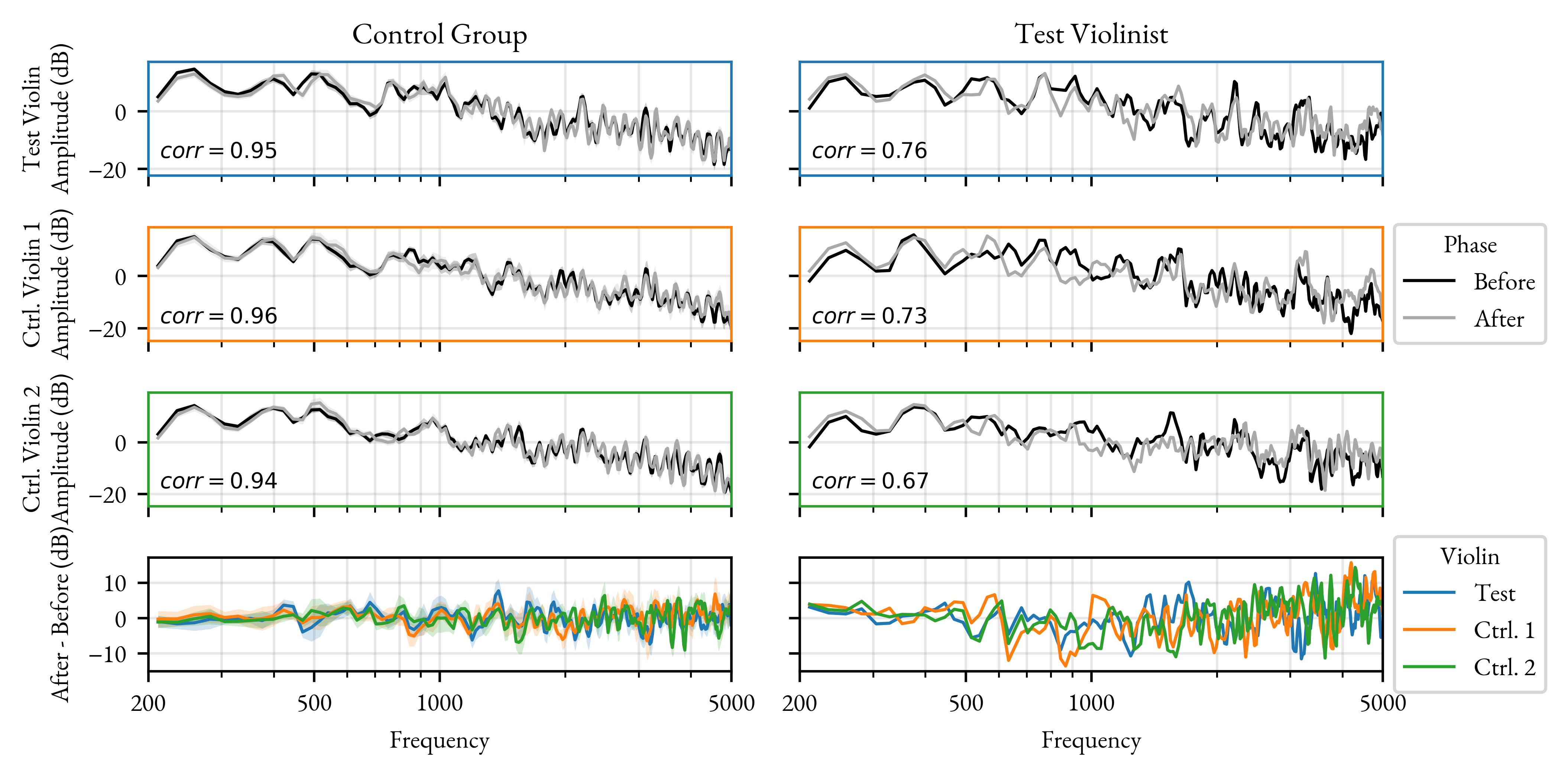}
\caption{Long-Term Average Spectra (LTAS) of scales recorded on the Test Violin and the two Control Violins. The left column displays the average across the Control Group (10 violinists), while the right column displays the Test Violinist. Black curves indicate the ``Before'' phase and gray curves indicate the ``After'' phase; the Pearson correlation coefficient between these two states is provided in the bottom-left corner of each panel. The bottom panel displays the mean LTAS magnitude difference between the ``Before'' and the ``After'' phases. For the Control Group, shaded areas represent the 95\% confidence interval.}\label{fig:recordings}
\end{figure*}

Figure \ref{fig:recordings} presents the LTAS of scales recorded on the Test Violin and the two Control Violins. The data are organized by cohort: the Control Group (left column) and the Test Violinist (right column). The top three rows show the LTAS for each instrument, while the bottom row shows the difference between the two phases. The spectral analysis confirmed that no significant drift exceeding measurement uncertainty occurred in the high-frequency register ($>$ 5~kHz). Consequently, the plots are restricted to the 200--5000~Hz range, which, as noted in \cite{ballesteros2025}, captures the primary discriminative spectral features of each instrument.

The left column of Figure \ref{fig:recordings} displays the average spectra of the Control Group (N=10). Since these musicians did not participate in the daily playing intervention, their recordings serve as independent probes of the instrument's physical state. Visually, the ``After'' curves (gray) for the Test Violin superimpose closely onto the ``Before'' curves (black), and this quantitatively supported by the high Pearson correlation coefficients computed between the two phases. Crucially, the small spectral difference for the Test Violin is of the same order of magnitude as that for the Control Violins, as shown by the highly comparable Pearson correlation coefficients for all three instruments. This confirms that the Test Violin did not undergo any acoustical evolution based on the recordings of a diverse panel of players.

The right column of Figure \ref{fig:recordings} displays the spectra of the Test Violinist. Although spectral differences between the ``Before'' and ``After'' phases appear larger than for the Control Group---likely due to the absence of averaging and thus to a sampling effect---they remain highly comparable across all three violins, as evidenced by their correlation coefficients. This suggests that the Test Violinist did not modify their sound production strategy to ``optimize'' this particular instrument, nor did they significantly adapt their playing technique in response to the six months of practice.

\subsection{Bowing parameters}
\begin{figure*}
\centering
\includegraphics[]{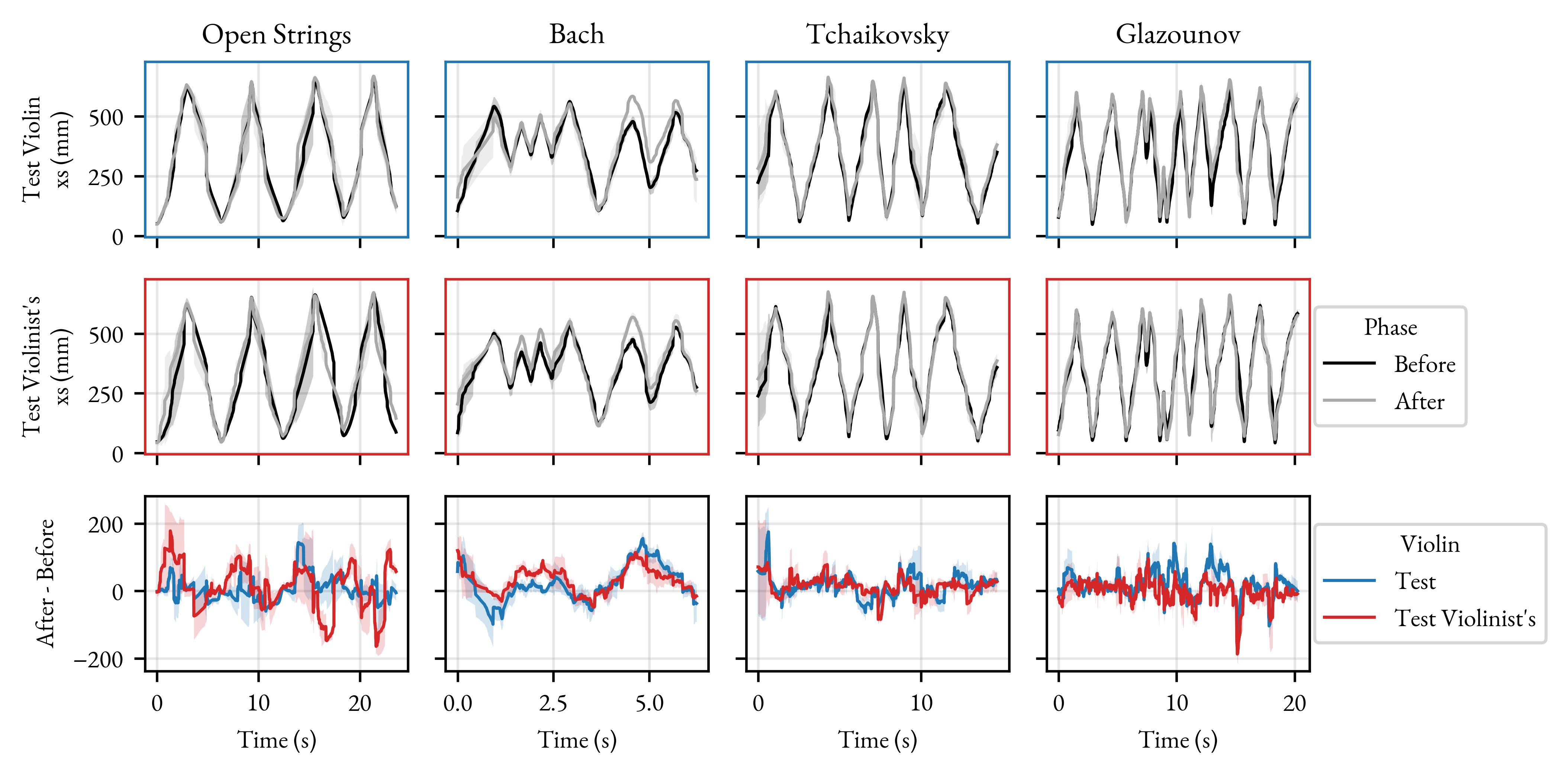}
\caption{Aligned bow position profiles (\(x_{s}\)) recorded on the Test Violin and the Test Violinist's Personal Violin, during playing by the Test Violinist. Columns correspond to distinct musical excerpts. Black curves denote the ``Before'' phase and gray curves denote the ``After'' phase. The bottom panel displays the mean bow position difference between the ``After'' and the ``Before'' phases. Shaded bands indicate the 95\% confidence interval across repeated takes ($N=3$ for the ``Before'' phase, $N=2$ for the ``After'' phase).}\label{fig:mocap_xs}
\end{figure*}

\begin{figure*}
\centering
\includegraphics[]{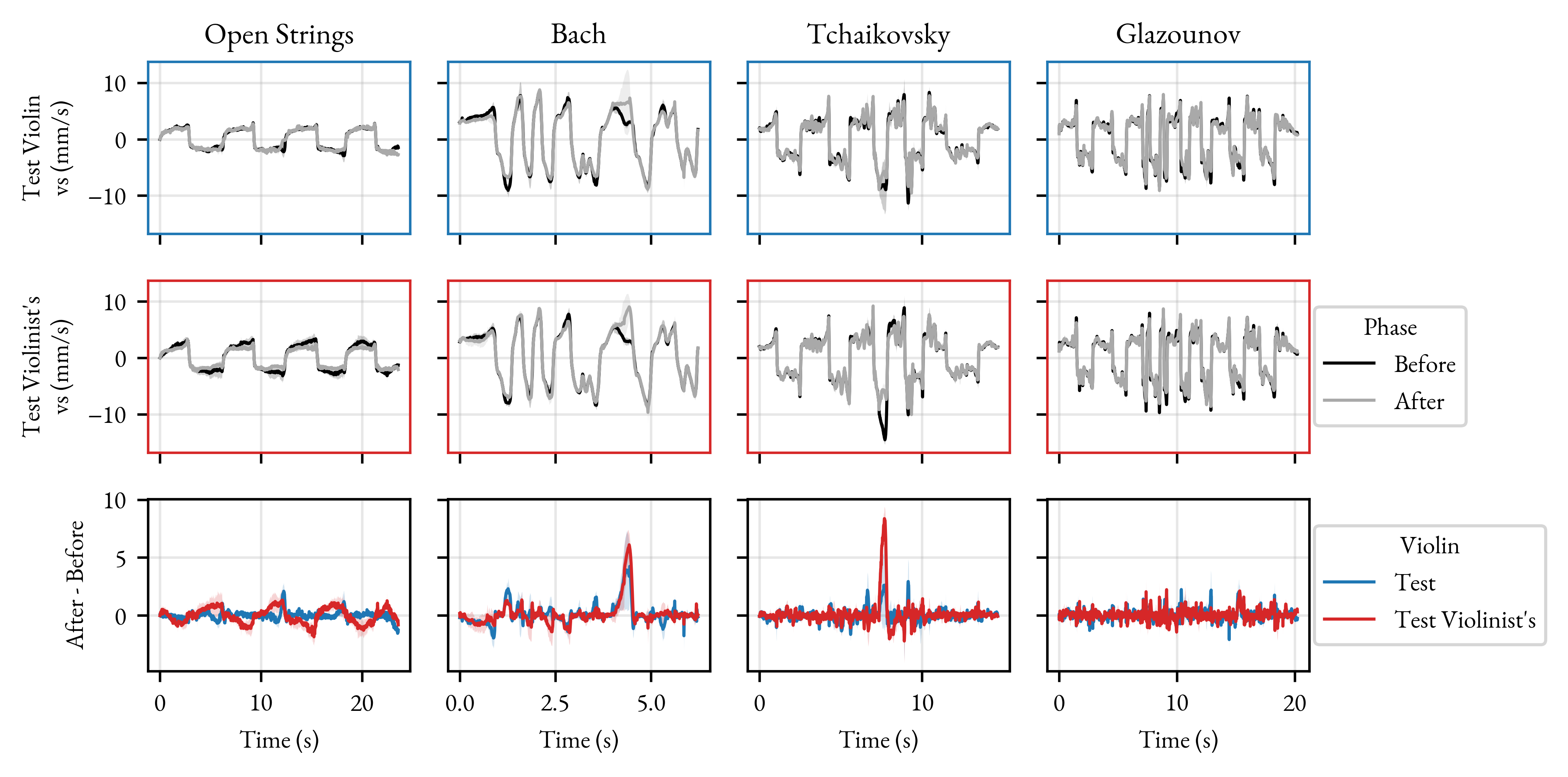}
\caption{As Fig. 5, but for bow velocity ($v_s$).}\label{fig:mocap_vs}
\end{figure*}

\begin{figure*}
\centering
\includegraphics[]{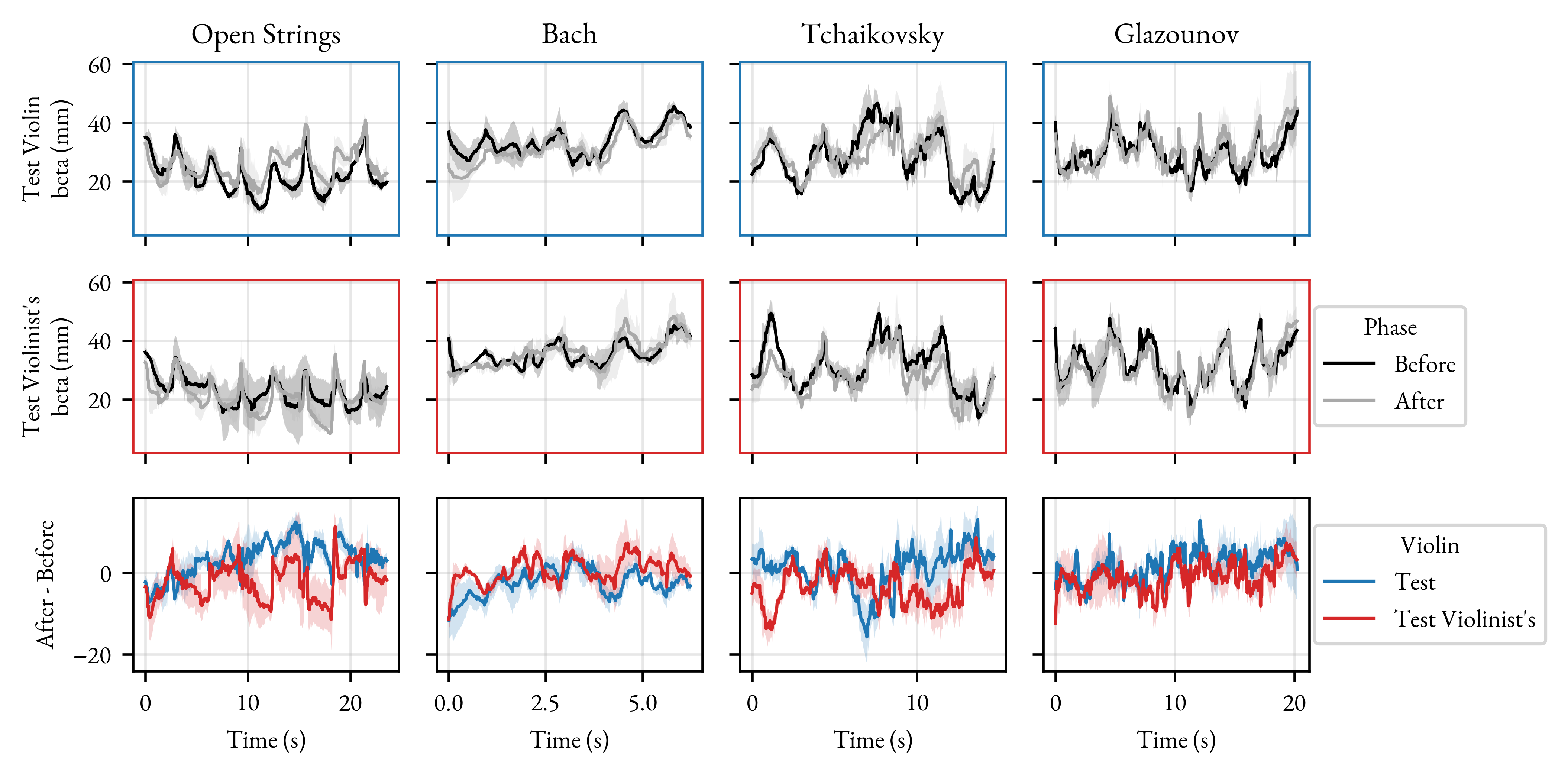}
\caption{As Fig. 5, but for bow-bridge distance ($beta$).}\label{fig:mocap_beta}
\end{figure*}

\begin{figure*}
\centering
\includegraphics[]{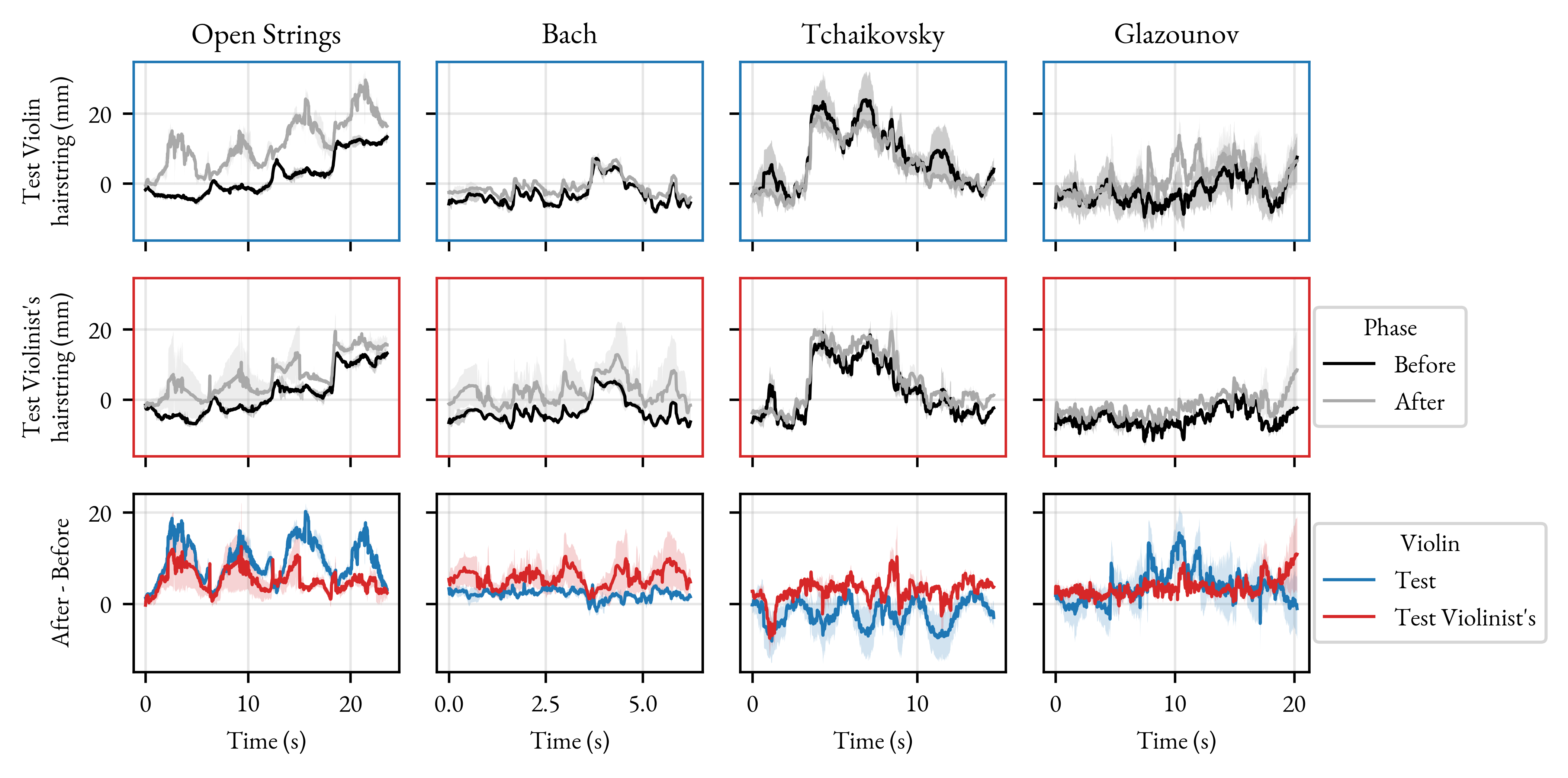}
\caption{As Fig. 5, but for bow-string distance \(hairstring\).}\label{fig:mocap_hairstring}
\end{figure*}

Figures \ref{fig:mocap_xs}, \ref{fig:mocap_vs}, \ref{fig:mocap_beta}, and \ref{fig:mocap_hairstring} present respectively the aligned bow position, bow velocity, bow-bridge distance and bow-string distance profiles for the Test Violin and the Test Violinist's Personal Violin. Each column corresponds to a distinct musical excerpt. Black curves indicate the ``Before'' phase and gray curves indicate the ``After'' phase. Visually, the Test Violinist exhibits high stability in their bowing strategy for each excerpt. The ``After'' curves (gray) often superimpose onto the ``Before'' curves (black) for both instruments.

Inspection of the bowing parameter differences reveals that the deviations for the Test Violin are of the same order of magnitude as those observed for the Personal Violin. The changes in bow parameters for the Test Violin do not exceed the baseline variability established by the familiar instrument. This suggests that the measured variations are attributable to natural motor variability or measurement reproducibility rather than a specific behavioral adaptation to the Test Violin.

\section{Discussion and Conclusions}\label{sec:d&c}
The primary objective of this longitudinal study was to disentangle the physical and behavioral components of the violin ``playing-in'' phenomenon. Our results provide a robust, double-null finding. First, regarding the instrument, the Test Violin exhibited no measurable drift in input admittance or radiated spectral energy (LTAS) that exceeded the environmental variability defined by the Control Violins. Second, regarding the player, the Test Violinist displayed remarkable kinematic stability; their bowing strategy (velocity profiles, contact point) was statistically indistinguishable between the ``Before'' and the ``After'' phases.

These findings align with recent rigorous investigations in musical acoustics \cite{inta2005, piacsek2024b}, reinforcing the view that the violin is a mechanically stable system over the timescale of months. The stability of the admittance curves supports material science reviews \cite{obataya2017} suggesting that vibration-induced changes in wood are either negligible under normal playing conditions or reversible due to environmental fluctuations. Furthermore, the lack of macroscopic motor adaptation suggests that expert technique is robust and ``top-down.'' The soloist appeared to impose a consistent motor signature upon the instrument rather than plastically reorganizing their technique to accommodate the violin's specific feedback. This double null result---neither the violin nor the player changed---suggests that the widely reported experience of ``playing-in'' may not be rooted in physical or motor reality.

These findings challenge the traditional narrative of instrument selection. If high-quality violins are physically stable and do not induce significant motor adaptation over six months, musicians and buyers should trust their initial assessment of an instrument. Identifying a violin as ``having potential to open up'' may be a misleading heuristic.

While this study improves upon previous designs by including control groups, it is limited by the sample size of the test condition (\(N = 1\) played violin, \(N = 1\) test player). Although the null result is robust for this specific dyad, we cannot categorically rule out that other instruments or players might exhibit different behaviors. 

Additionally, the relationship of differences in bridge mobility to perceptual differences is not understood. Therefore, we cannot exclude that some of the small changes measured in the bridge mobilities, even if they are of the same physical order for the Test Violin as for the Control Violins, might be of perceptual importance. To address this, perceptual evaluations from the 11 players were also collected. Their analysis may shed additional light on possible perceptual effects. Furthermore, the sound recordings will be used for blinded listening tests to explore whether listeners can hear differences between ``Before'' and ``After''. The results will be presented in a subsequent paper.

If perceptual differences are indeed observed on the Test Violin, a more granular investigation of the Motion Capture data may be warranted. The current analysis focused on macroscopic bowing parameters (\(v_{s}, x_{s}, \beta, \text{hair-string}\)); it remains possible that motor adaptation occurs at a micro-temporal level below the resolution of this study.

Conversely, if no perceptual differences are found, the ``Mere Exposure Effect''---where repeated exposure and familiarity are cognitively processed as an improvement in quality---could offer a more plausible explanation for the persistent belief in the playing-in phenomenon \cite{peretz1998}.



\acknowtext

The authors would like to thank the participants in the experiment. We also extend our gratitude to the Conservatoire National Supérieur de Musique et de Danse de Paris (CNSMDP) for providing the instruments and facilities. We thank Julien Dubois and Léa Martinez for their help organizing the study. Special thanks to Mehdi Maglach, François Longo, Emma Barthe, and Victor Salvador Castrillo for their assistance during the measurement sessions.

\conflict
The authors declare no conflict of interest. 

\dataavailability
The violin acoustics dataset generated during this study, including bridge admittance, motion capture (MoCap) trajectories, and audio recordings, is available in the Zenodo repository at \url{https://doi.org/10.5281/zenodo.18696785}  \cite{paugetballesteros2026}. The Python code used to access and process the data is available at \url{https://github.com/hugo-paugesteros/CNSM-dataset}.

\bibliographystyle{unsrtnat}
\bibliography{bib}

\end{document}